\pdfoutput = 1
\documentclass[journal]{IEEEtran}
\usepackage{cite}
\usepackage{amsmath,amssymb,amsfonts}
\usepackage{graphicx}
\usepackage{textcomp}
\usepackage{comment}
\usepackage{multirow}
\usepackage{algorithm}
\ifCLASSOPTIONcompsoc
\usepackage[caption=false, font=normalsize, labelfont=sf, textfont=sf]{subfig}
\else
\usepackage[caption=false, font=footnotesize]{subfig}
\usepackage{algpseudocode}
\usepackage{amsmath}
\usepackage{epsfig}
\usepackage{float}

\ifCLASSINFOpdf
\else
\fi

\begin{document}
%
\title{Resource Allocation Using Gradient Boosting Aided Deep Q-Network for IoT in C-RANs}
%
%
%

\author{Yifan~Luo,~Jiawei~Yang,~Wei~Xu,~\IEEEmembership{Senior~Member,~IEEE,}~Kezhi~Wang,~\IEEEmembership{Member,~IEEE},\\
and~Marco~Di~Renzo,~\IEEEmembership{Senior~Member,~IEEE} 
 
\thanks{Y. Luo and J. Yang made the same contribution to this paper.}
\thanks{Y. Luo and W. Xu are with the National Mobile Communications Research Laboratory (NCRL), Southeast University, Nanjing 210096, China (e-mail: roey.luoyif@gmail.com; wxu@seu.edu.cn).}
\thanks{J. Yang is with the Department of Computer Science and Engineering, Southeast University, Nanjing 210096, China (e-mail: Jiawei-Young@outlook.com).}
\thanks{K. Wang is with the Department of Computer and Information Sciences, Northumbria University, Newcastle upon Tyne NE1 8ST, U.K.}
\thanks{M. Di Renzo is with the Laboratoire des Signaux et Syst\`emes, CNRS, CentraleSup\'elec, Univ Paris Sud, Universit\'e Paris-Saclay, 3 rue Joliot Curie, Plateau de Saclay, 91192 Gif-sur-Yvette, France (e-mail: marco.direnzo@centralesupelec.fr).}}

%



\maketitle

\begin{abstract}
In this paper, we investigate dynamic resource allocation (DRA) problems for  Internet of Things (IoT) in real-time cloud radio access networks (C-RANs), by combining gradient boosting approximation and deep reinforcement learning to solve the following two major problems. 
Firstly, in C-RANs, the decision making process of resource allocation is time-consuming and computational-expensive, motivating us to use an approximation method, i.e. the gradient boosting decision tree (GBDT) to approximate the solutions of second order cone programming (SOCP) problem. 
Moreover, considering the innumerable states in real-time C-RAN systems, we employ a deep reinforcement learning framework, i.e., deep Q-network (DQN) to generate a robust policy that controls the status of remote radio heads (RRHs).
We propose a GBDT-based DQN framework for the DRA problem, where the heavy computation to solve SOCP problems is cut down and great power consumption is saved in the whole C-RAN system.
We demonstrate that the generated policy is error-tolerant even the gradient boosting regression may not be strictly subject to the constraints of the original problem. Comparisons between the proposed method and existing baseline methods confirm the advantages of our method.
\end{abstract}


%
\IEEEpeerreviewmaketitle

\section{Introduction}
%
%
%
%

\IEEEPARstart{T}{he} requirement and development of Internet of Things (IoT) services, a key challenge in 5G, have been continuously rising, with the expanding diversity and density of IoT devices\cite{b0}. Cloud radio access networks (C-RANs)\cite{b1} are regarded as the promising mobile network architecture to meet this new challenge. Specifically, C-RANs separate base stations into radio units, which are commonly referred as remote radio heads (RRHs), and signal processing centralized baseband unit (BBU) Pool. In a C-RAN, BBU can be placed in a convenient and easily accessible place, and RRHs can be deployed up on poles or rooftops on demand.  It is expected that C-RAN architecture will be an integral part of future deployments to enable efficient IoT services.

Dynamic resource allocation (DRA) for IoT in C-RANs is indispensable to maintain acceptable performance. In order to get the optimal allocation strategy, several works have tried to apply convex optimizations, like second order cone programming (SOCP) in \cite{b16}, semi-definite programming (SDP) in \cite{b5} and mix-integer programming (MIP) in \cite{bbbb24}. However, in real-time C-RANs where the environment keeps changing, the efficiency of the above methods in finding the optimal decision faces great challenges. Attempts have been made in reinforcement learning (RL) to increase the efficiency of the solution procedure in \cite{b16} \cite{bbbb8}.
 
RL has shown its great advantages to solve DRA problems in wireless communication systems and for IoT. Existing methods to DRA problem in RANs generally model it as a RL problem \cite{bbbb8}\cite{bbbbb8}\cite{bbbb7}, by setting different parameters as the reward. For instance, the work in \cite{bbbb8} regarded the successful transmission probability of the user requests as the reward, and another work in \cite{bbbb7} set the sum of average quality of service (QoS) and averaged resource utilization of the slice as the reward. However, with the increase of the complexity in allocation problems, the search space of solutions tends to be infinite, which is hard to be tackled. 


With the combination of RL and deep neural network (DNN) \cite{bbb0}, deep reinforcement learning (DRL) has been proposed and applied to address the above problems in \cite{DD1}\cite{DD2}\cite{DD3}. By utilizing the ability of extracting useful features directly from the high-dimensional state space of DNN, DRL is able to perform end-to-end RL \cite{bbb0}. With the assistance of DNN, problems of large search space and continuous states are no longer the insurmountable challenges. 

%
 
To apply DRL framework in DRA problems, the design of reward, action and state becomes vital. The action set needs to be enumerable in most circumstances. The work in \cite{b16} used a two-step decision framework to guarantee its enumerability, by changing the state of one RRH at each epoch, which performs well in the models with innumerable states. 

Furthermore, in DRA problems, how to get optimal allocation strategy will be finally turned into another optimization problem in most cases, i.e., convex optimization problem \cite{b2}, which can be solved mathematically. 
Unfortunately, traditional algorithms\cite{bbbb9}\cite{bbbb10}\cite{bbbb11} for solving the convex optimization problem, such as SOCP still faces significant limitations, such as time-consuming, making it hard to generate a policy for large-scale systems.


Recent works have achieved significant improvement in computational efficiency by applying the DNN approximator \cite{bbbb13}\cite{bbbb14}\cite{CC11}\cite{CC22} to DRA problems. However, the unstable performance of DNN in regression process makes it hard to achieve good performance \cite{bbbb22}. With a large number of hyper-parameters, fine tuning becomes even harder in practical system. Some researchers discussed and investigated this problem in computability theory and information theory domains, e.g., in \cite{bbbb12}. 

Gradient boosting machine (GBM) \cite{bbbb15} is one member of boosting algorithms family\cite{bbbb19}\cite{bbbb20}, a sub-branch of ensemble learning\cite{bbbb16}\cite{bbbb17}\cite{bbbb18}. 
It has been firmly established as one of state-of-the-art approaches in machine learning (ML) community, and it has played a dominating role in existing data mining and machine learning competitions\cite{bbbb21} due to its fast training and excellent performance. However, to the best of our knowledge, few works applied this method to the DRA problem, even to other regression problems in communication systems. 

In this paper, to efficiently address DRA problem for IoT in C-RANs with innumerable states, one common form of DRL, namely the deep Q-network (DQN) is employed. Moreover, to tackle the difficulties in obtaining the reward in DQN in low latency, a tree-based GBM, i.e., gradient boosting decision tree (GBDT) is utilized to approximate the solutions of SOCP. Then, we demonstrate the improvement of our method by comparing it to the traditional methods under simulations.

The main contributions of this paper are as follows:
\begin{itemize}
	\item We first give the model of dynamic resource allocation problem for IoT in the real-time C-RAN. Then, we propose a GBDT-based regressor to approximate the SOCP solution of the optimal transmitting power consumption, which serves as the immediate reward needed in DQN. By doing so, there is no need to solve the original SOCP problem every time, and therefore great computational cost can be saved.
	
	\item Next, we aggregate the GBDT-based regressor with a DQN to propose a new framework, where the immediate reward is obtained from GBDT-based regressor instead of SOCP solutions, to generate the optimal policy to control the states of RRHs. The proposed framework can save the power consumption of the whole C-RAN system for IoT.
	
	\item We show the performance gain and complexity reduction of our proposed solution by comparing it with the existing methods.
\end{itemize}

The remainder of this paper is organized as follows. Section \uppercase\expandafter{\romannumeral2} presents the related works, whereas system model is given in 
Section \uppercase\expandafter{\romannumeral3}. Section \uppercase\expandafter{\romannumeral4} introduces the proposed GBDT-based DQN framework. The simulation results are reported in Section \uppercase\expandafter{\romannumeral5}, followed by the conclusions presented in Section \uppercase\expandafter{\romannumeral6}.




\section{Related Works}



The resource allocation problem under C-RANs is normally interpreted into an optimization problem, where one needs to search the decision space to find an optimal combinatorial set of decisions to optimize different goals \cite{b2} \cite{b3} \cite{b4} based on current situations. Although numerous researchers devoted their time in finding solutions to optimization problems, most of them are still hard or impossible to be tackled with traditional pure mathematical methods. RL has been recently applied to address those problems.

In \cite{CC0}, a model-free RL model was adopted to solve the adaptive selection problem between backhaul and fronthaul transfer modes, which aimed to minimize the long-term delivery latency in fog radio access network (F-RAN). Specifically, an online on-policy value-based strategy State-Action-Reward-State-Action (SARSA) with linear approximation was applied in this system. Moreover, some works have proposed more efficient RL methods to overcome slow convergence and scalability issues in traditional RL-based algorithms, such as Q-learning. In \cite{CC1}, four methods, i.e. state space reduction techniques, convergence speed up methods, demand forecasting combined with RL algorithm and DNN were proposed to handle the aforementioned problems, especially to deal with the huge state space.

Furthermore, as reported in \cite{ab1}, DQN achieved a better performance on resource allocation problems, compared with the traditional Q-learning based method. 
In practice, the size of possible state space may be very large or even infinite, which makes it impossible to traverse each state that required by the traditional Q-learning. Approximation methods can address this kind of problem that they maps the continuous and innumerable state space to a near-optimal Q-value space in consecutive setting, rather than Q-table. DNN shows its advantage of approximation in the high-dimensional space in many domains.
Therefore, the adoption of DNN to estimate Q-value can improve the system performance and computing efficiency, as reported in the simulation results from \cite{ab1}.

In \cite{b16}, a two-step decision framework was adopted to solve the enumerability problem of action space in C-RANs. The DRL agent first determined which RRH to turn on or turn off, and then the agent got the resource allocation solution by solving a convex optimization problem. Any other complex actions can be decomposed into the two-step decision, reducing the action space significantly. Moreover, the work in \cite{bbbb8} shows the impractical use of SA (i.e., Single BS Association) scheme even in a small-scale C-RAN. Specifically, SA scheme abandoned the collaboration of each RRH and only supported few users. This research is a guidance to our research.

The works in \cite{bbbb8} and \cite{bbbb7} all adopted the DRL method to solve resource allocation problems in the RAN settings. In \cite{bbbb7}, the concept of intelligent allocation based on DRL was proposed to tackle the cache resource optimization problem in F-RAN. To satisfy user's QoS, the caching schemes should be intelligent, i.e. more effective and self-adaptive. Considering the limitation of cache space, this requirement challenges the design of schemes, and it motivates the adoption of DRL technique.

As reported in \cite{bbbb8}, a DRL-based framework is used in more complicated resource allocation problems, i.e., virtualized radio access networks. Based on the average QoS utility and resource utilization of users, the DQN-based autonomous resource management framework can make virtual operations to customize their own utility function and objective function based on different requirements.


In this paper, to improve the system efficiency, we propose a novel gradient-boosting-based DQN framework for resource allocation problem, which significantly improves the system performance through offline training and online running.

To the best of our knowledge, there is few works to apply gradient boosting machine to approximate solutions of convex optimization problems in wireless communication and we are the first to propose this framework.

\section{System Model}
\subsection{Network Model}

\begin{figure}
\centering
\includegraphics[width=3.3 in]{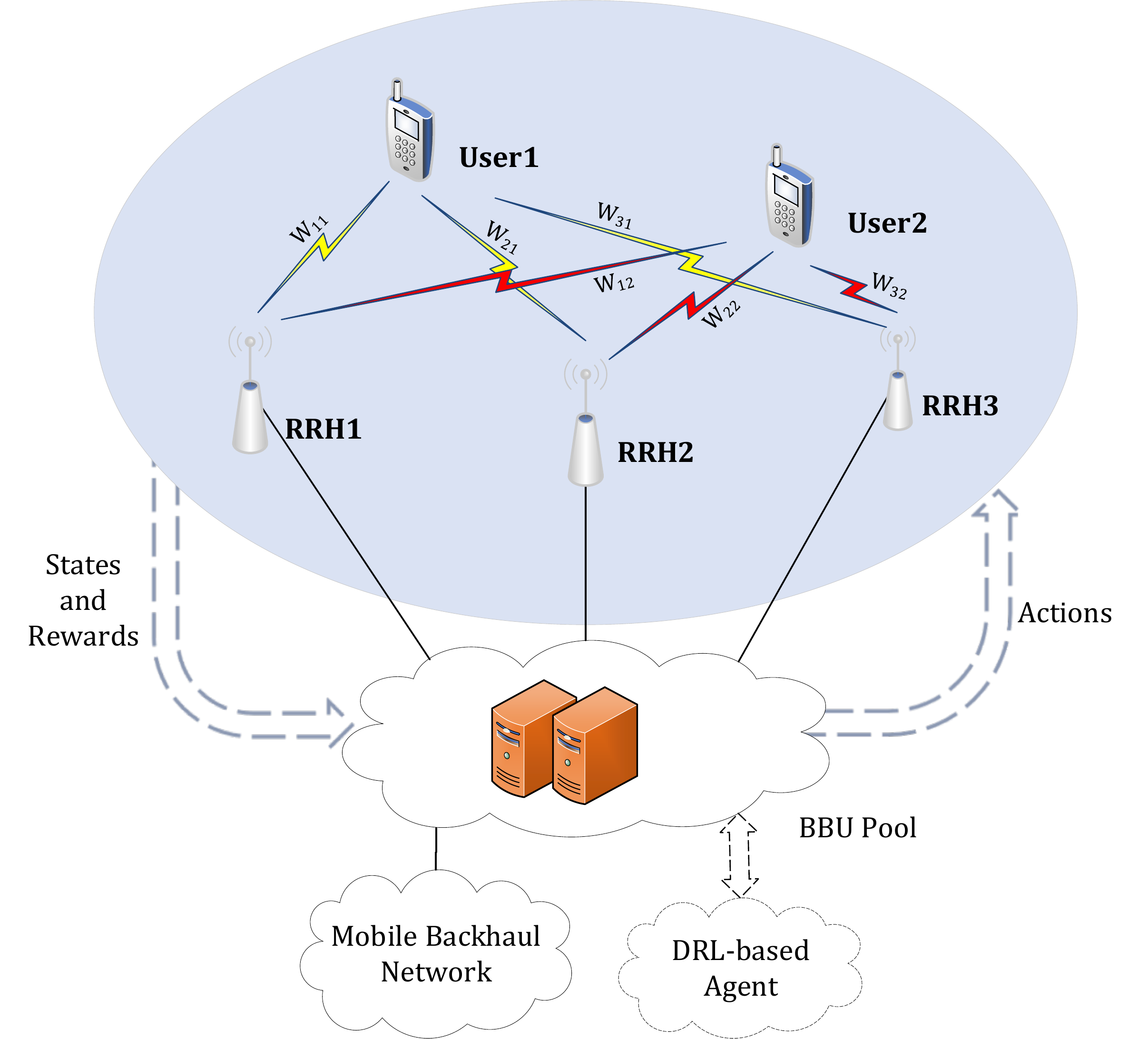}
\caption{Dynamic Resource Allocation for IoT under DRL framework in C-RANs.}
\label{fig1}
\end{figure}

We consider a typical C-RAN architecture where there is a single cell model with a set of $m$ RRHs denoted by $\mathcal{R} =\{r_1,r_2,...,r_m\}$ and a set of $n$ users which can be some IoT devices denoted by $\mathcal{U}=\{u_1,u_2,...,u_n\}$. In the DRA for IoT in C-RAN as shown in Fig.~\ref{fig1}, we can get the current states, i.e. the state of each RRHs and the demands of IoT device users, from the networks in $K$-th decision epoch $t_k$. All the RRHs are connected to the centralized BBU pool, meaning all information can be shared and processed by the DQN-based agent to make decisions, i.e. turning on or off the RRHs. We simplify the model by making assumption that all RRHs and users are equipped with a single antenna, which is readily to be generalized into the multi-antenna case by using technique proposed in \cite{b17}. 

Then, the corresponding signal-to-interference-plus-noise ratio (SINR) at the receiver of $i$-th user $u_i$ can be given as:
\begin{equation}
{SINR}_{i}=\frac{\left|\textbf{h}_{u_i}^T \textbf{w}_{u_i}\right|^2}{\sum_{u_j\neq u_i}{\left|\textbf{h}_{u_i}^T\ \textbf{w}_{u_j}\right|^2+\sigma^2}},\ u_i\in\mathcal{U}
\label{eq3_1_1}
\tag{1}
\end{equation}
where $\textbf{h}_{u_i}=[h_{r_{1}u_{i}},h_{r_{2}u_{i}},...,h_{r_{u}u_{i}}]^T$ denotes the channel gain vector and each element $h_{r_l u_i}$ denotes the channel gain from RRH $r_l\in\mathcal{R}$ to user $u_i$; $\textbf{w}_{u_i}=[w_{r_1 u_i},w_{r_2 u_i},...,w_{r_m u_i}]^T$ denotes the vector of all RRHs beamforming to user $u_i$ and each element $ w_{r_l u_i}$ denotes the weight of beamforming vector in RRH $r_l\in\mathcal{R}$ distributed to user $u_i$ and $\sigma^2$ is the noise. 

According to the Shannon formula, the data rate of user $u_i$ can be given as:
\begin{equation}
R_{i}=B\log_2{\left(1+\frac{{SINR}_{i}}{\mathrm{\Gamma}_m}\right)},\ u_i\in\mathcal{U}
\label{eq3_1_2} 
\tag{2}
\end{equation}
where $B$ is the channel bandwidth and $\mathrm{\Gamma_m}$ is the SINR margin depending on a couple of practical considerations, e.g., the modulation scheme.

The relationship of the transmitting power and the power consumed by the base station can be approximated to be nearly linear, according to \cite{b19}. Then, we apply the linear power model for each RRH as:
\begin{equation}
P_{i}=\left\{
\begin{split}
&P_{r_i,A}+\frac{1}{\eta}P_{r_i,T} & & {r_i\in\mathcal{A}}\\
&P_{r_i,S} & & {r_i\in\mathcal{S}}
\end{split} 
\right.
\label{eq3_1_3}
\tag{3}
\end{equation}
where $P_{r_i,T}=\sum_{u_j\in\mathcal{U}} |w_{r_i u_j}|^2$ is the transmitting power of RRH $r_i$; $\eta$ is a constant denoting the drain efficiency of the power amplifier; and $P_{r_i,A}$ is the power consumption of RRH $r_i$ when $r_i$ is active without transmitting signals. In the case of no need for transmission, $r_i$ can be set to the sleep mode, whose power can be given by $P_{r_i,S}$. Thus, one has $\mathcal{A}\cup\mathcal{S}=\mathcal{R}$.

In addition, we take consideration of the power consumption for the state transition of RRHs, i.e. the power consumed to change RRHs' states. We put the RRHs which reverse states in the current epoch to the set $\mathcal{T}$ and use $P_{r_i,T}$ to denote the power to change the mode between $Active$ and $Sleep$, i.e. we assume they share the same power consumption.
Therefore, in the current epoch, the total power consumption of all RRHs can be written as:
\begin{equation}
\begin{split}
P_\mathrm{Total}=&{\underbrace{\sum_{r_i\in\mathcal{A}}\sum_{u_j\in\mathcal{U}}{\frac{1}{\eta}\left|w_{r_i u_j}\right|^2}}_{\text{Transmitting Power}}}+\underbrace{\sum_{r_i\in\mathcal{A}} P_{r_i, A}+\sum_{r_i\in\mathcal{S}} P_{r_i, S}}_{\text{State Power}}\\
+&{\underbrace{\sum_{r_i\in\mathcal{T}} P_{r_i, T}}_{\text{Transition Power}}}.
\end{split}
\label{eq3_1_4}
\tag{4}
\end{equation} 

\subsection{CP-beamforming}

From Equation \eqref{eq3_1_4}, one can see that the latter two parts are easy to be calculated, which are composed by some constants and only relying on the current state and action. To minimize $P_\mathrm{Total}$, it is necessary for us to calculate the minimal transmitting power in each epoch, which depends on the allocation scheme of beamforming weights in active RRHs.
Therefore, this optimization problem can be expressed as:\\
$\bullet$ \ \textbf{Control Plane (CP)-Beamforming:}
\begin{alignat}{2}
\min \limits_{w_{r_i u_j}} \quad & P_\mathrm{T} & \tag{5} \label{eq3_2_1} \\
\mathrm{s.t.} \quad & P_\mathrm{T}=\sum_{r_i\in\mathcal{A}}\sum_{u_j\in\mathcal{U}}\left|w_{r_i u_j}\right|^2 & \tag{5.1} \label{eq3_2_2}\\
& R_{i} \le B\log_2{\left(1+\frac{{SINR}_{i}}{\mathrm{\Gamma m}}\right)}, \ u_i\in\mathcal{U} & \tag{5.2} \label{eq3_2_3} \\
& \sum_{u_j\in\mathcal{U}}\left|w_{r_i u_j}\right|^2\le\ P_{r_i}, \ r_i\in\mathcal{A} & \tag{5.3} \label{eq3_2_4}
\end{alignat}
where the objective is to get the minimal total transmitting power given the states of RRHs and user demands. Also,
the variables $w_{r_i u_j}$ are distributive weights corresponding to beamforming power; $R_{i}$ is defined as the user demand; ${SINR}_{i}$ is given by Equation \eqref{eq3_1_1} and $P_{r_i}$ is the transmitting power constraint for RRH $r_i$. Also, Constraint \eqref{eq3_2_3} ensures the demand of all users will be met, whereas Constraint \eqref{eq3_2_4} ensures the limitation of transmitting power in each RRH. 

As shown in \cite{b5}, the above CP-beamforming can be transformed into a SOCP problem. Therefore, we rewrite the above optimizations as:\\
$\bullet$ \ \textbf{Modified CP-Beamforming:}
\begin{alignat}{2}
\min \limits_{w_{r_i u_j}} \quad & P_\mathrm{T} & \tag{6} \label{eq3_2_5} \\
\mathrm{s.t.} \quad & \sum_{u_j\in\mathcal{U}}\left|\textbf{h}_{u_i}^H \textbf{w}_{u_j}\right|^2 +\sigma^2\le \mu_{u_i}\left|\textbf{h}_{u_i}^H \textbf{w}_{u_i}\right|^2, \ u_i\in\mathcal{U} & \tag{6.1} \label{eq3_2_6}\\
 & \sum_{u_j\in\mathcal{U}}\left|w_{r_i u_j}\right|^2\le\ P_{r_i}, \ r_i\in\mathcal{A} & \tag{6.2} \label{eq3_2_7}\\
 & \sum_{r_i\in\mathcal{A}}\sum_{u_j\in\mathcal{U}}\left|w_{r_i u_j}\right|^2\le P_\mathrm{T} & \tag{6.3} \label{eq3_2_8}\\
 & \mu_{i}=\frac{\iota_{i}+1}{\iota_{i}}, \ u_i\in\mathcal{U}& \tag{6.4} \label{eq3_2_9} \\
 & \iota_{i}=\mathrm{\Gamma}_m(2^\frac{R_{i}}{B}-1) & \tag{6.5} \label{eq3_2_10}
\end{alignat}
where we apply variable $P_\mathrm{T}$ to replace the optimization \eqref{eq3_2_2} by adding Constraint \eqref{eq3_2_8}, which is a common method in transformation process \cite{b22}. We also rewrite Constraint \eqref{eq3_2_3} as Constraint \eqref{eq3_2_6} and apply some simple manipulations to get the above modified optimization.

Now, it is ready to see that the above Modified CP-Beamforming optimization is the same as a standard SOCP problem. By using the iterative algorithm mentioned proposed in \cite{b21}, we can get the optimal solutions.
It is worth noting that the CP-Beamforming optimization may have no feasible solutions. In this case, it means more RRHs should be activated to satisfy the user demands. In this case, we will give a large negative reward to the DQN agent and jump out of the current training loop.

Then, we can calculate the total power consumption by applying Equation \eqref{eq3_1_4}. In the following part, we propose the DQN-based framework to predict the states of RRHs and adopt GBDT to approximate the solutions of the aforementioned SOCP problems.


\begin{figure*}
\centering
\includegraphics[width=7 in]{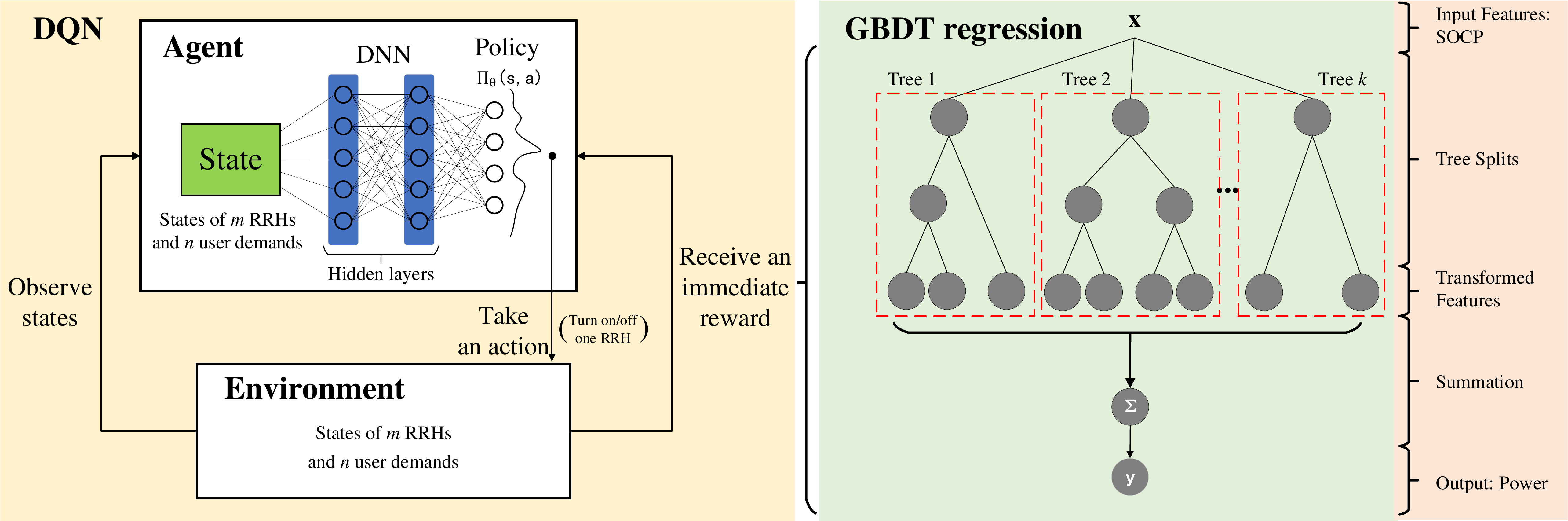}
\caption{The DQN-based scheme for dynamic resource allocation.}
\label{DQN-based}
\end{figure*}

\section{GBDT Aided Deep Q-Network for DRA in C-RANs}
\subsection{State, Action Space and Reward Function}
Our goal in the aforementioned DRA problem is to generate a policy that minimizes the system's power consumption at any state by taking the best action. Here, the best action refers to the action that contributes the least to overall power consumption in a long term but also satisfies user demands, system requirements and constraints among all the available actions. The fundamental idea of RL-based method is to abstract an agent and an environment from the given problem to generate the environment model \cite{b26} and employ the agent to find the optimal action in each state, so as to maximize the cumulative discounted reward by exploring the environment and receiving immediate reward signalled by the environment.

To apply RL method in our problem, we transform the system model defined in Section \uppercase\expandafter{\romannumeral3} into a RL model. The general assumption that future reward is discounted by a factor of $\gamma$ per time-step is made here. Then, the cumulative discounted reward from time-step $t$ can be expressed as:
\begin{equation}
    R_t =\mathbb{E}\left[\sum_{k=t}^{\infty}{\gamma^k r_k(s_k,a_k) \arrowvert{s_t=s,a_t=a}}\right]
\label{eq3_3_1}
\tag{7.1}
\end{equation}
where $\mathbb{E}(\cdot)$ denotes mathematical expectation; $r_k(\cdot)$ denotes the $k$-th reward; $s_k$ denotes the $k$-th state and $\gamma \in (0,1]$ denotes the discount factor. If $\gamma$ tends to 0, the agent only considers the immediate reward; whereas if $\gamma$ tends to 1, the agent focuses on the future reward. Moreover, the infinity over the summation sign indicates the endless sequence in DRA problem.

Leveraging the common definition in Q-learning, the optimal action-value function $Q^*(s,a)$ is defined as the greatest mathematical expected cumulative discounted reward reached by taking action $a$ in state $s$ and then following a subsequently optimal policy, which guarantees the optimality of cumulative future reward. The function $Q^*(s,a)$ strongly follows the \textit{Bellman equation}, a well-known identity in optimality theory. In this model, the optimal action-value function $Q(s,a)$ to represent the maximum cumulative reward from state $s$ with action $a$ can be expressed as:
\begin{equation}
    Q^{*}(s,a)=\mathbb{E}\left[{r_{(s,a)}+\gamma \max \limits_{a'}Q^{*}(s',a')}\arrowvert {s,a}\right]
\label{eq3_3_2}
\tag{7.2}
\end{equation}
where $r_{(s,a)}$ denotes the immediate reward received at state $s$ if action $a$ is taken; $a'$ denotes the possible action in the next state $s'$, and other symbols are of the same meaning as Equation \eqref{eq3_3_1}. The expression means that the agent takes action $a$ in the state $s$, receiving the immediate reward $r_{(s,a)}$, and then subsequently follows an optimal trajectory that leads to greatest $Q(s,a)$ value.

In a general view, $Q^*(s,a)$ demonstrates how promising the final expected cumulative reward will be if action $a$ is taken in state $s$ in a quantitative way. That is to say, in DRA problem, how much power consumption the C-RAN can cut down if it decides to take the action $a$, i.e switches on or off one selected RRH when observing the state $s$, i.e. a set of user demands and the states (i.e. sleep/active) of RRHs. Since the true value of $Q^*(s,a)$ can never be known, our goal is to employ DNN to learn an approximation $Q(s,a)$. For the following sections, $Q(s,a)$ just denotes the approximated  $Q^*(s,a)$ and has all the same properties of $Q(s,a)$. 

The generic policy function $\pi(s)$ defined in the context of RL is used here, which can be expressed as:
\begin{equation}
    \pi(s)=\arg\max \limits_{a} Q^{*}(s,a)
\label{eq3_3_3}
\tag{7.3}
\end{equation}
where $\pi(s)$ is the \textit{argmax} of the action-value function $Q(s,a)$ over all possible actions in a specific state $s$. The policy function leads to the action that maximize the $Q(s,a)$ values in all states.

The state, action and reward defined in our problem are given as:
\begin{itemize}
\item \textbf{State: }
The state has two components that one is a set of states of RRHs and the other is a set of demands from users. Specifically, $\mathcal{Y}=[y_1,y_2,...,y_m]$ is defined as the set of all $m$ RRHs' states, in which $y_i\in\{0,1\}$ denotes the state of RRH $i$. In the case of $y_i=0$, RRH $i$ is in the sleep state, whereas $y_i=1$ means that it is in the active state. $\mathcal{D}=[d_1,d_2,...,d_n]$ is defined as the set of all $n$ users' demands, and $d_j\in[d_{\mathrm{min}},d_{\mathrm{max}}]$ denotes the demand of user $j$, in which $d_{\mathrm{min}}$ is the minimum of all demands and $d_{\mathrm{max}}$ is the maximal demand. Thus, the state of RL is expressed as $\mathcal{D}\cup\mathcal{Y}=[y_1,y_2,...,y_m,d_1,d_2,...,d_n]$ and its cardinality is $m+n$.
\item \textbf{Action: }
In each decision epoch, we enable the RL agent to determine the next state of one RRH. We use a set of $ A =\{\alpha_1,\alpha_2,\cdots,\alpha_m\}$ to denote the action space, in which $\alpha_i \in \{ 0,1\},i=1,2,\cdots,m$. If $\alpha_i=1$, it means RRH $i$ changes the state, otherwise the RRH remains its current state in next epoch. Then, the action space can be substantially reduced. It is noteworthy that we set the constriction that $0 \leqslant \sum_{i=1}^{m}{\alpha_i} \leqslant 1$, which means only one or none of all RRH states will alter its state and reduces the space into the size of $m+1$.
\item \textbf{Reward: }
To minimize the total power consumption, we define the immediate reward as the difference between the upper bound of power consumption. The actual power consumption is expressed as:
\begin{equation*}
\mathcal{r}_k = P_\mathrm{UB}-P_\mathrm{total}   
\end{equation*}
where $P_\mathrm{UB}$ denotes the upper bound of the power consumption obtained from the system setting, and $P_\mathrm{total}$ denotes the actual total power consumption of the system that is composed of three parts defined in Equation \eqref{eq3_1_4}. To be more specific, the reward is defined to minimize the system power consumption under the condition of satisfying the user demands, which requires us to solve the optimization problem according to Equation \eqref{eq3_2_5}, shown in Section \uppercase\expandafter{\romannumeral3}.
\end{itemize}

To sum up, the policy mentioned in this work is a function that maps the current state $s$, the set of user demand and RRHs status, to the best action $a$, turning on or off one RRH, that minimizes the overall power consumption of the whole system. 
\subsection{Gradient Boosting Decision Tree}
GBM is a gradient boosting framework that can be applied to any classifiers or regressors. To be more specific, GBM is the aggregation of base estimators (i.e., classifiers or regressors) that any base estimators like $K$ nearest neighbor, neural network and naive Bayesian estimators can be fitted into the GBM. Better base estimators advocate higher performance. Among all kinds of GBM, a prominent one is based on decision tree, called gradient boosting decision tree (GBDT), which has been gaining its popularity for years due to its competitive performance in different areas. In our framework, the GBDT is applied to the regression task due to its prominent performance. 

The concept of GBDT is to optimize the empirical risk via steepest gradient descent in hypothesis space by adding more base tree etimators. Considering the regression task in our work, given a dataset with $n$ entities of different states and their corresponding rewards generated by simulation and solving SOCP, one can have 
\begin{equation*}
D={(x_i,p_i)}(|D|=n,x_i\in D\cup\mathcal{Y}, p_i\in \mathbb{S})
\end{equation*}
where $x_i$ denotes the state representation  $[y_1^{i},y_2^{i},...,y_m^{i},\\d_1^{i},d_2^{i},...,d_n^{i}]$ of system model, whereas $p_i$ denotes the corresponding solution of SOCP solver from Equation \eqref{eq3_2_5}, in line with the definition of the \textit{Reward} function. To optimize the empirical risk of regression is to minimize the expectation of a well-defined loss function over the given dataset $D$, which can be express as:
\begin{equation*}
    \mathbb{E}_{D}[{L(\widehat{P},P)}]=l(\widehat{P},P)+\Omega(\phi) = l(f(X),P)+\Omega(\phi)
\tag{8}
\end{equation*}
where $\phi$ denotes the model itself and $f(X)$ is the final mapping to approximate $P$, which is our fitting object, the power comsumption. $X$ is the set of $x_i$ representing system model, and $P$ is the set of $p_i$ representing solution of SOCP solver. Here the first term is model prediction loss, which is a differentiable convex function to measure the distance between true power consumption and estimated power consumption; and $L_2$ loss (i.e., mean-square error) is applied in this task. The latter term is the regularization penalty applied to constrain model complexity, contributing to finalize a model with less over-fitting and better generalization performance. 

The choice of prediction loss and regularization penalty alters circumstantially. Also, the penalty function is given by:
\begin{align*}
    \Omega(f)=\beta T+\frac{1}{2}\lambda{\|w\|}^2    
\end{align*}
where $\beta$ and $\lambda$ are two hyper-parameters, while $T$ and $w$ are the numbers of trees ensembled and weights owned by each tree, respectively. When the regularization parameter is set to zero, the loss function falls back to the traditional gradient tree boosting method \cite{bbbb25}.

In GBDT, it starts with a weak model that simply predicts the mean value of $P$ at each leaf and improves the prediction by aggregating $K$ additive fixed size decision trees as base estimators to predict the pseudo-residuals of previous results. The final prediction is linear combination of all the output from $K$ regression trees. The final estimator function as adverted in \eqref{eq16} can be expressed as follow:
\begin{equation}
    f(x) = \sum_{k=0}^{K} f_k (x) = f_0 (x) + \sum_{k=0}^{K} \theta_k \phi_k (x)
\label{eq16}
\tag{9}
\end{equation}
where $f_0$ is the initial guess, $\phi_k (x)$ is the base estimator at the iteration $k$ and $\theta_k$ is the weight for the $k_{th}$ estimator or a fixed learning rate. The product $\theta_k \phi_k (x)$ denotes the step at iteration $k$.

\subsection{GBDT-based Deep Q-Network (DQN)}
In this section, we will show how to apply GBDT-based DQN scheme to solve our DRA problem for IoT in real-time C-RAN, by using the previously defined states, actions and reward. Traditional RL methods, like Q-learning, compute and store the Q value for each state-action group into a table. It is unrealistic to apply those methods in our problem, as the state-action groups are countless and the demands of users in a state are continuous variables. Therefore, DQN is considered to be best solutions for this problem. Similar with the related works, e.g. \cite{bbbb7}\cite{ab1}, we also apply experience replay buffer and fixed Q-targets in this work to estimate the action-value function $Q\left(s,a\right)$. 


In our framework, two stages are included, i.e., offline training and online decision making as well as regular training: 

\begin{itemize}
\item  For offline training stage, we pre-train DQN to estimate the value of taking each action in any specific states. To achieve this, millions of system data are generated in terms of all RRHs’ states, user demands and its corresponding system power consumption by simulation and solving SOCP problem given in equation \eqref{eq3_2_5}. Then, the GBDT is employed to estimate the immediate reward to alleviate the expensive computation in solving the SOCP problem for further training and tuning.

\item  For online decision making and regular tuning, we load the pre-trained DQN to generate the best action to take for our proposed DRA problem in real-time. This is achieved by employing the policy function defined in \eqref{eq3_3_3}, which maximizes the $Q(s,a)$ in state $s$. To emphasize, the $Q(s,a)$ function tells how much the system can cut down the power consumption if it decides to take the action $a$ when seeing the state $s$. Then, the DQN observes the immediate reward $r_t$ obtained from GBDT approximation and observes next state $s_{t+1}$. In an online regular tuning scheme, the DQN will not immediately update model parameters when observing new states but to store the new observations to memory buffer. Then, under some given conditions, the DQN will fine-tune its parameters according to that buffer. This allows DQN to dynamically adapt to new patterns regularly.
\end{itemize}

The whole algorithm is given in Algorithm 1, whereas the framework of GBDT-based DQN is given by Fig.~\ref{DQN-based}. The $\theta$ denotes the set of model parameters. The loss function is  $L_2$ loss (i.e., mean-square error), which indicates the difference between Q target and model output. \textbf{S$i$} refers to the step $i$ in Algorithm 1.

\begin{algorithm}
\caption{ GBDT-based DQN framework }
\begin{algorithmic}
\State{\textbf{Offline:}}
\State{\textbf{S1:} Generate millions of data randomly via SOCP solver;}
\State{\textbf{S2:} Pre-train GBDT with those data and save its model;}
\State{\textbf{S3:} Pre-train DQN with those data from \textbf{S6} to \textbf{S15}, and}
\State{\qquad then save the well-trained network and its correspond-}
\State{\qquad ing experience memory $D$;}

\State{\textbf{Online:}}
\State{\textbf{S4:} Load replay memory with capacity $N$ from offline-}
\State{\qquad trained experience memory $D$ and load GBDT model;}
\State{\textbf{S5:} Set action-value function $Q$ with weights $\theta$ from offline-}
\State{\qquad trained network;}
\State{\textbf{S6:} \textbf{For} \text{each episode} $t$-th}
    \State{\textbf{S7:} \quad \ \textbf{a)} Offline: with probability $\varepsilon$ select a random action} 
    \State{\quad \quad \quad $a_t$, otherwise select $a_t =\pi(s)$ when observing a}
    \State{\quad \quad \quad new state $s$;}
    \State{\quad \quad \quad \textbf{b)} Online: directly select $a_t =\pi(s)$ when observing} 
    \State{\quad \quad \quad a new state $s$;}
    \State{\textbf{S8:} \quad \ Execute action $a_t$;}
    \State{\textbf{S9:} \quad \ Obtain reward $r_t$ from \textbf{a)} Offline: SOCP solver or} 
    \State{\quad \quad \quad \textbf{b)} Online: GBDT machine, and observe $s_{t+1}$;}
    \State{\textbf{S10:} \quad  Store transition $(s_t,a_t,r_t,s_{t+1})$ in $D$;}
    \State{\textbf{S11:} \quad  \textbf{if} \textbf{a)} Offine or \textbf{b)} Online: reach the given condition} 
    \State{\quad \quad \quad \ \textbf{then} execute from \textbf{S12} to \textbf{S14};}
    \State{\textbf{S12:} \qquad \quad Sample random mini-batch of transitions} 
    \State{\qquad \qquad \quad $(s_j,a_j,r_j,s_{j+1})$ from $D$;}
    \State{\textbf{S13:} \qquad \quad Set}
     \State{\qquad \qquad \quad $ y_{j}=\left\{
    \begin{aligned}
    &r_j, \ \ \text{if episode terminates at step } j \text{+1}\\
    &r_j+\gamma \max \limits_{a'} Q(\phi_{j+1},a';\theta^{-}),\ \text{otherwise} 
    \end{aligned}
    \right.
    $}
    \State{\textbf{S14:} \qquad \quad Apply the gradient descent step on}
    \State{\qquad \qquad \quad ${(y_i-Q(\phi_{j},a_j;\theta))}^2$ with respect to the net-}
    \State{\qquad \qquad \quad work parameters $\theta$;}
\State{\textbf{S15: EndFor}}
\label{DQN11}
\end{algorithmic}
\end{algorithm}

In Fig.~\ref{DQN-based}, one can see that the left side describes a DQN framework, illustrating the agent, the environment and how to get the reward. Specifically, the agent will observe a new state from the environment after taking an action and then it will receive an immediate reward signalled by the reward function from GBDT approximator.
Traditional DQN obtains the reward by solving the SOCP optimization, which can not be real-time, as explained before. In our architecture, we adopt GBDT regression (i.e., the right side of Fig.~\ref{DQN-based}) to obtain the reward, which can operate in a online process in real-time. 

We also give the training process of GBDT in the Appendix.



\subsection{Error Tolerance Examination (ETE)}
Our target is to use GBDT to approximate the typical SOCP problem in C-RANs under the framework of DQN. Thus, it is important to evaluate its practical performance. 
The error from GBDT or DNN will influence the optimality of the given scheme, even worsening the performance of whole system power consumption. Therefore, the examination of error influence is of vital significance. Considering its important role in the whole DRA problem, we emphasize the concept of error tolerance examination (ETE) here. Specifically, in the simulation, we will first compare the result of the optimal decision provided by CP-Beamforming solution with the near-optimal decision from GBDT or DNN approximation solution, and then evaluate its performance in the dynamic resource allocation settings.

\section{Simulation Results}

In this section, we present the simulation settings and performance of the proposed GBDT-based DQN solutions. 
We take the definition of channel fading mode from previous work as \cite{bbbb23}:
\begin{equation}
h_{r,u}=10^{\frac{-L(d_{r,u})}{20}}\sqrt{\varphi_{r,u} s_{r,u}}\mathcal{G}_{r,u} 
\label{eq4_1} 
\tag{4.1}
\end{equation}
where $L(d_{r,u})$ is the path loss with the distance of $d_{r,u}$; $\varphi_{r,u}$
is the antenna gain; $s_{r,u}$ is the shadowing coefficient and $\mathcal{G}_{r,u}$
is the small-scale fading coefficient. The simulation settings are summarized in Table I. 

All training and testing processes are conducted in the environment equipped with 8GB RAM, Intel core i7-6700HQ (2.6GHz), python 3.5.6, tensorflow 1.13.1 and lightGBM 2.2.3.

\begin{table}  
\begin{center}  
\caption{Simulation Settings}  
\begin{tabular}{|p{25pt}|p{75pt}|p{110pt}|}  
\hline  
Symbol & Parameters & Value \\  
\hline  
$B$ & Channel bandwidth & 10 MHz  \\  
\hline  
$P_{r,max}$ & Max transmit power  &  1.0 W \\  
\hline  
$P_{r,A}$ & Active power & 6.8 W \\  
\hline  
$P_{r,S}$ & Sleep power & 4.3 W  \\  
\hline  
$P_{r,T}$ & Transition power & 2.0 W \\  
\hline  
$\sigma^2$ & Background noise & -102 dBm \\
\hline  
$\varphi_{r,u}$ & Antenna gain & 9 dBi  \\  
\hline  
$s_{r,u}$  & Log-normal shadowing &  8 dB \\  
\hline  
$\mathcal{G}_{r,u}$ & Rayleigh small-scale fading & $\mathcal{CN}(0, I)$ \\  
\hline  
$d_{r,u}$ & Path loss with a distance of (km) & $148.1 + 37.6 \mathrm{log_2} d_{r,u}$ dB \\  
\hline  
$d_{r,u}$ & Distance & Uniformly distributed in $[0,800]$ m \\  
\hline  
$\eta_l$ & Power amplifier efficiency & 25\% \\  
\hline
\multicolumn{3}{p{230pt}}{$^{\mathrm{a}}$W $=$ Watt, dB $=$ decibel, dBm $=$ decibel-milliwatts, dBi $=$ dB(isotropic).}
\end{tabular}  
\end{center}  
\end{table}

\begin{figure*}
    \centering
	  \subfloat[All RRHs open]{
       \includegraphics[width=2.3 in]{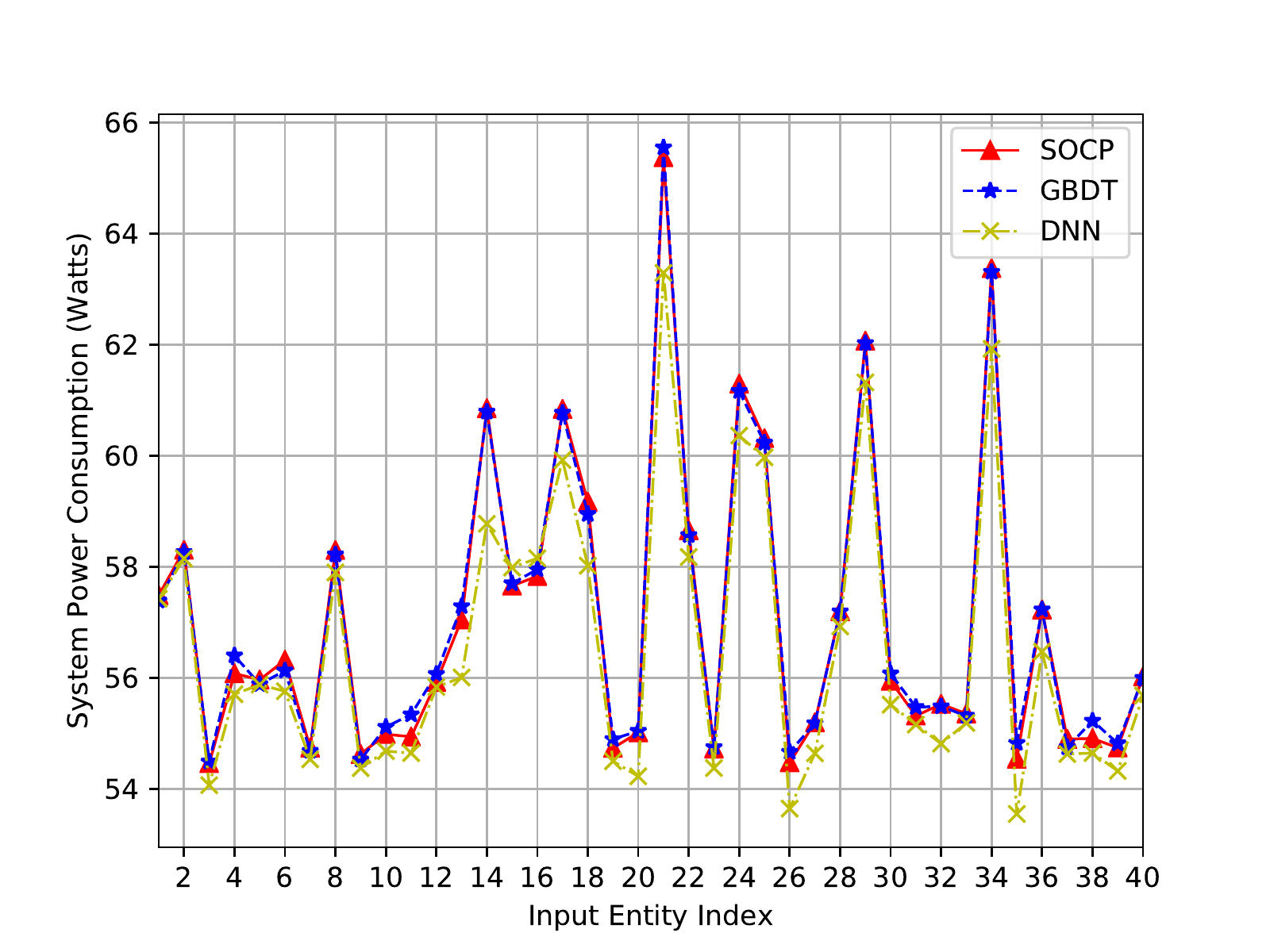}}
    \label{11a}\hfill
	  \subfloat[One RRH closed]{
        \includegraphics[width=2.3 in]{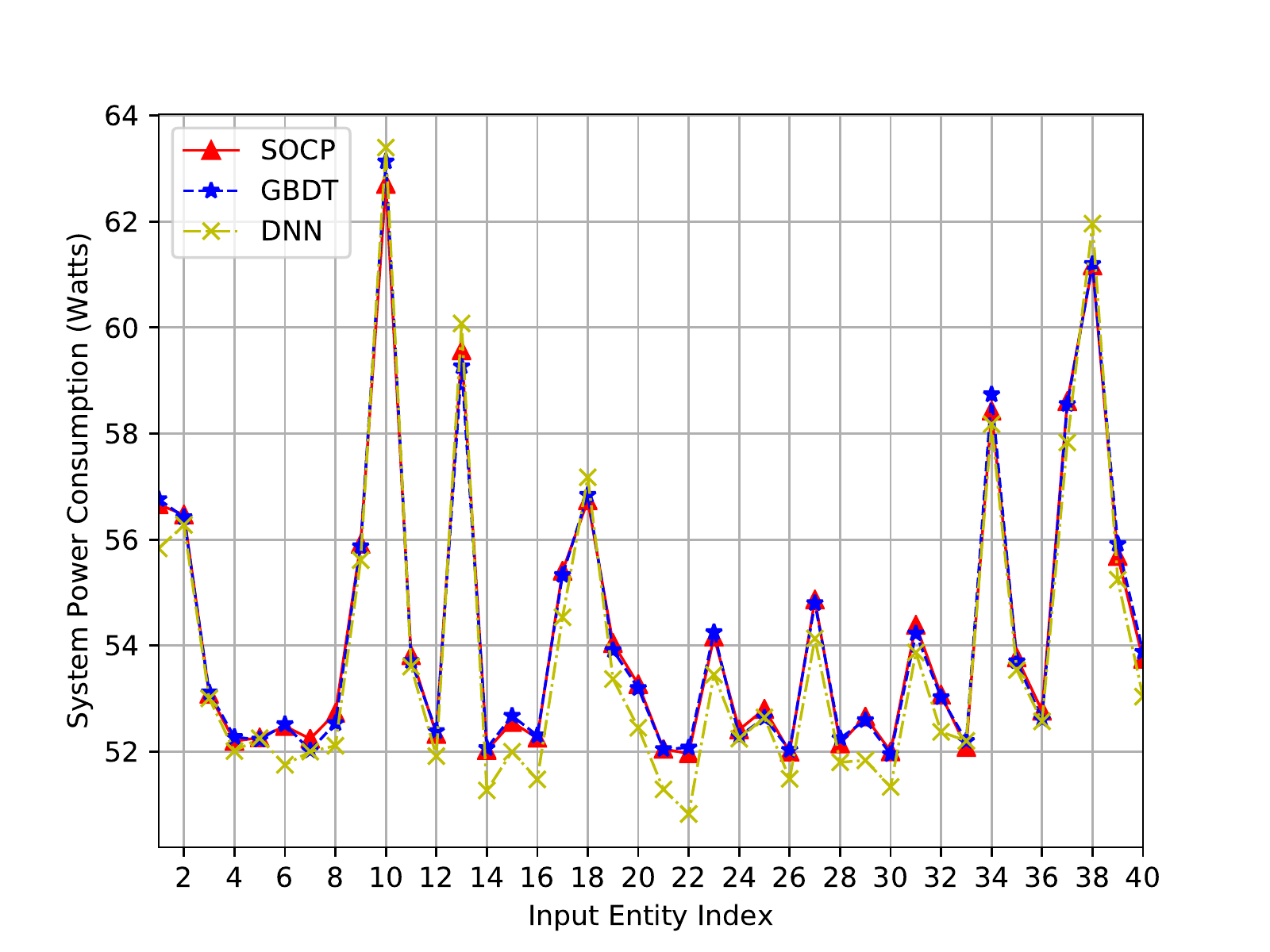}}
    \label{11b}\hfill
      \subfloat[Random states of RRHs]{
        \includegraphics[width=2.3 in]{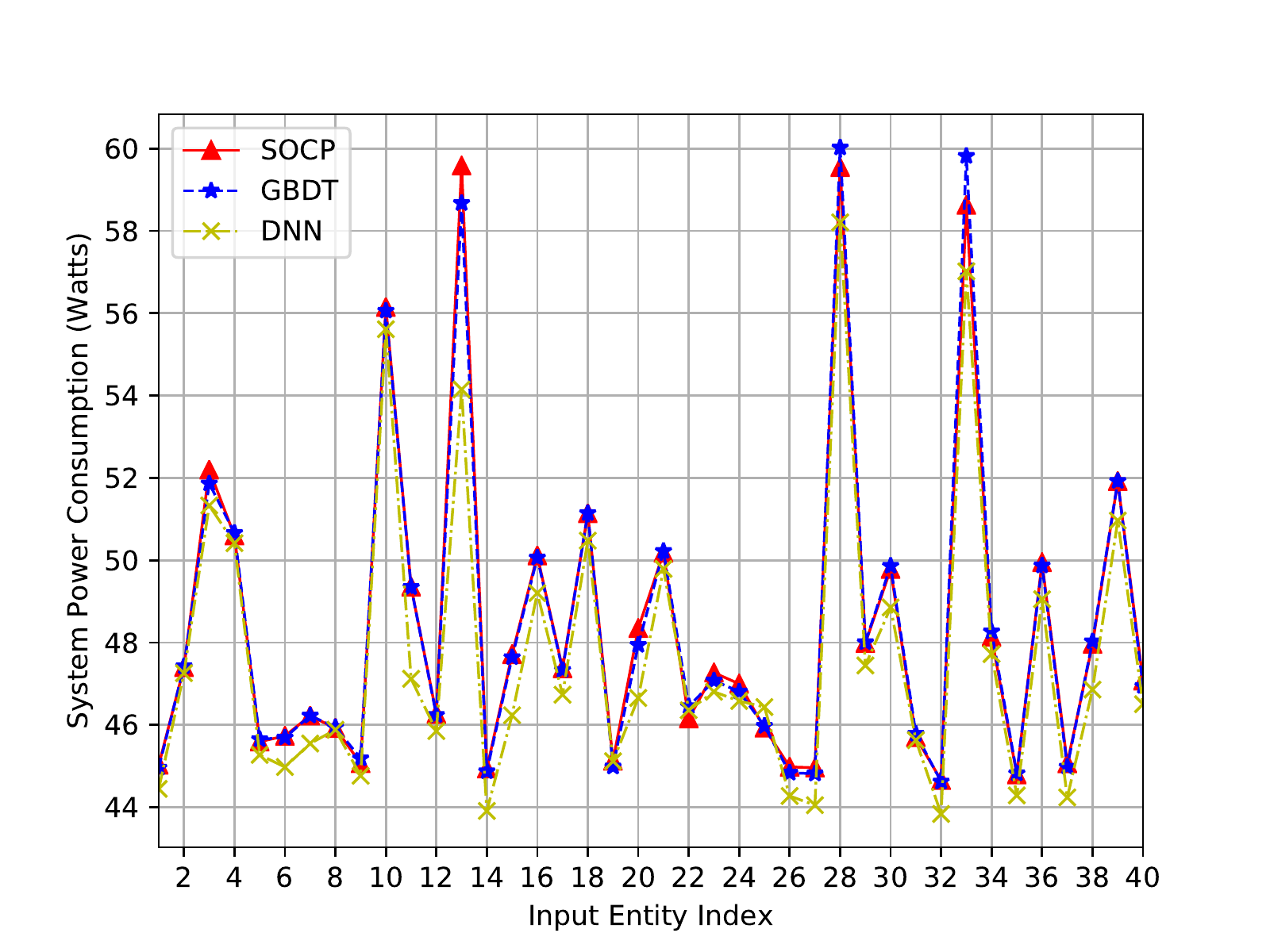}}
    \label{11c}
	  \caption{Approximation performance of GBDT approximator.}
	  \label{fig3-3} 
\end{figure*}

We compare our DQN-based solution containing GBDT approximator (abbreviated as DQN) with two other schemes: 

1) All RRHs Open (AO): all RHHs are turned on, which can serve each user; 

2) One RRH Closed (OC): one of those RHHs (chosen randomly) stays in the sleep state, which cannot serve any user. 

It is noteworthy that the in previous work \cite{bbbb26}, another solution in which only one random RRH is turned on, is also discussed in the dynamic resource allocation problem. However, it can hardly be applied to the practical systems \cite{b16}. Therefore, we do not compare it in this paper.

\subsection{GBDT-based SOCP Approximator}

\subsubsection{Computational Complexity}
We compare computational complexity between a GBDT approximator and solutions from traditional SOCP solver in \cite{CC1}. Firstly, a test set of 1000 entities are randomly generated in terms of status of RRHs and user demands. In addition, both the GBDT approximator and the traditional SOCP method are executed to predict or compute the outputs of that test set for 10000 times, respectively. One can see from Table II that GBDT approximator is much faster than SOCP solver, which prove the efficiency of GBDT approximator.

\begin{table}  
\begin{center}  
\caption{computational Complexity Comparison}  
\begin{tabular}{|p{70pt}|p{65pt}|p{65pt}|}  
\hline  
\multirow{2}{*}{System Input Setup}&\multicolumn{2}{|l|}{Average Time Per Input} \\  
\cline{2-3}  
& GBDT  & SOCP \\  
\hline  
6 RRHs and 3 users & 0.00079$s$ & 0.08281$s$ \\  
\hline  
8 RRHs and 4 users & 0.00077$s$ &  0.09387$s$ \\  
\hline  
12 RRHs and 6 users & 0.00070$s$ & 0.16240$s$ \\  
\hline  
18 RRHs and 9 users & 0.00075$s$ &  0.42803$s$ \\  
\hline   
\multicolumn{3}{p{230pt}}{$^{\mathrm{a}}$The time in above table is obtained by averaging 1000 different system inputs, each of which is recalculated by 10000 times through two algorithms respectively.}
\end{tabular}  
\end{center}  
\end{table}

\subsubsection{Fitting Property}
Then, we analyse the performance of GBDT approximator in specific situations, where we set that there are 8 RRHs and 4 users of IoT devices whose demands are ranging from 20\textit{Mbps} to 40\textit{Mbps} respectively. We compare it with DNN approximator. It applies the fully-connected net with 3 layers, each of which with 32, 64, 1 neurons respectively. Its activation function is a rectified linear unit (ReLU). Firstly, in Fig.~\ref{fig3-3}(a), we assume that all 8 RRHs are turned on. One can see from this figure that GBDT has better fitting performance than DNN. Then, we assume that there is one RRH switched off. One can see from Fig.~\ref{fig3-3}(b) that GBDT still fits very well with the SOCP solutions. In Fig.~\ref{fig3-3}(c), we assume that the states of all 8 RRH are set switched on or off randomly. As expected, GBDT has much better fitting performance, compared with the SOCP solutions.

\subsection{Training Effect of GBDT and DNN}
\begin{figure} 
    \centering
	  \subfloat[Training effect of GBDT within 4 seconds]{
       \includegraphics[width=3.5 in]{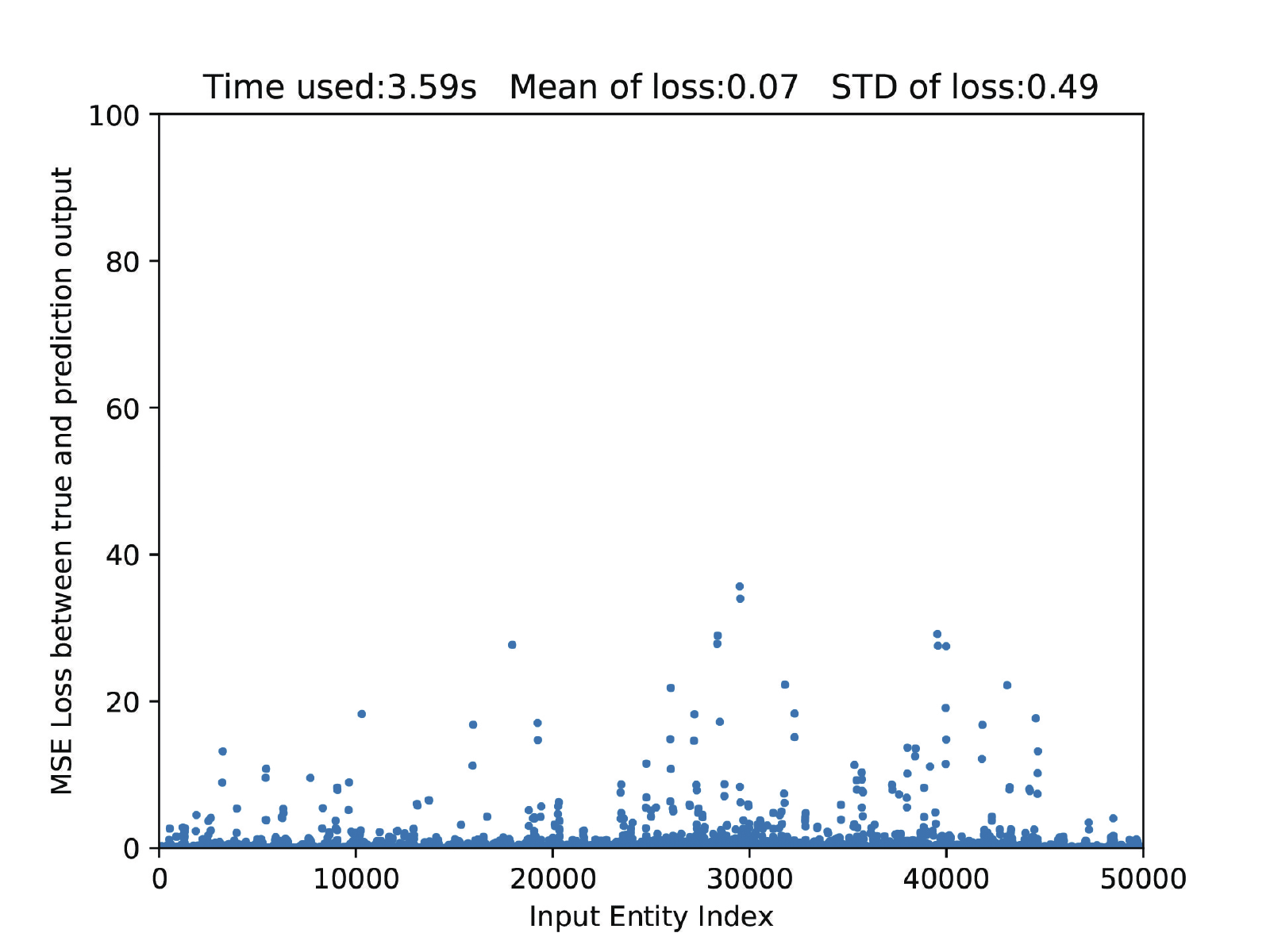}}
    \label{1a}\\
	  \subfloat[Training effect of DNN over 3 hours]{
        \includegraphics[width=3.5 in]{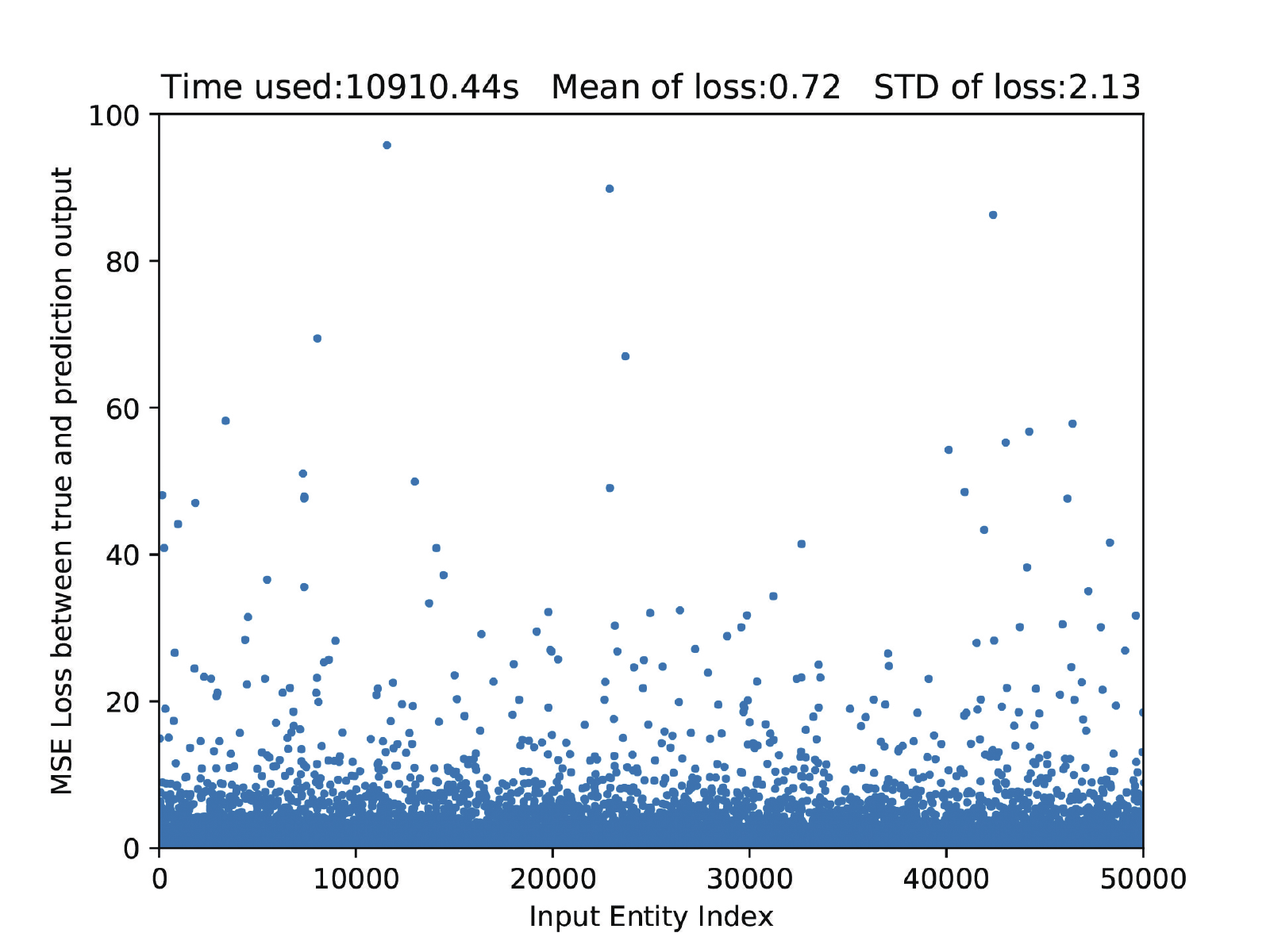}}
    \label{1b}
	  \caption{Training effect of GBDT and DNN.}
	  \label{fig3} 
\end{figure}

We demonstrate the training performance between the GBDT approximator and DNN aproximator by comparing the training effect in Fig.~\ref{fig3}. Mean squared error (MSE) is used here to calculate the loss. From Fig.~\ref{fig3}, one can see that even trained with far more time, the loss of DNN is still higher than that of GBDT. One also notices that GBDT has less parameters to adjust and therefore has quicker training process.


The specific comparison is not unfolded here, as it is not the focus of this paper. Next, we will examine the performance of GBDT-based DQN solutions. 

\subsection{System Performance}
In this section, we consider there are 8 RRHs and 4 users, whose demands are randomly selected. We change the user demands every 100 ms. The performance of AO, OC and GBDT-based DQN is compared next. 

\subsubsection{Instant Power}
\begin{figure} 
    \centering
	  \subfloat[Demands from 20\textit{Mbps} to 40\textit{Mbps}]{
       \includegraphics[width=3.5 in]{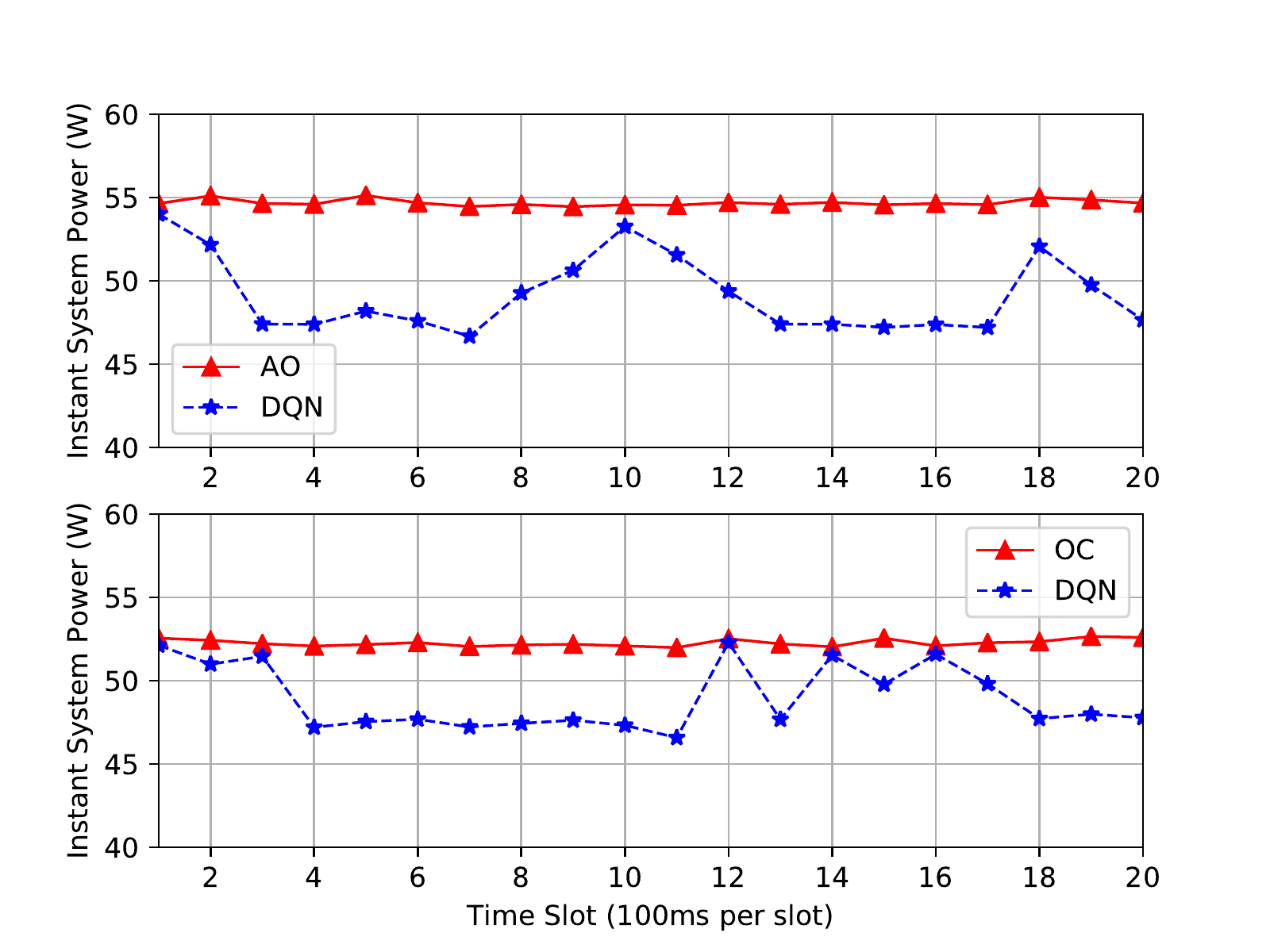}}
    \label{2a}\\
	  \subfloat[Demands from 20\textit{Mbps} to 60\textit{Mbps}]{
        \includegraphics[width=3.5 in]{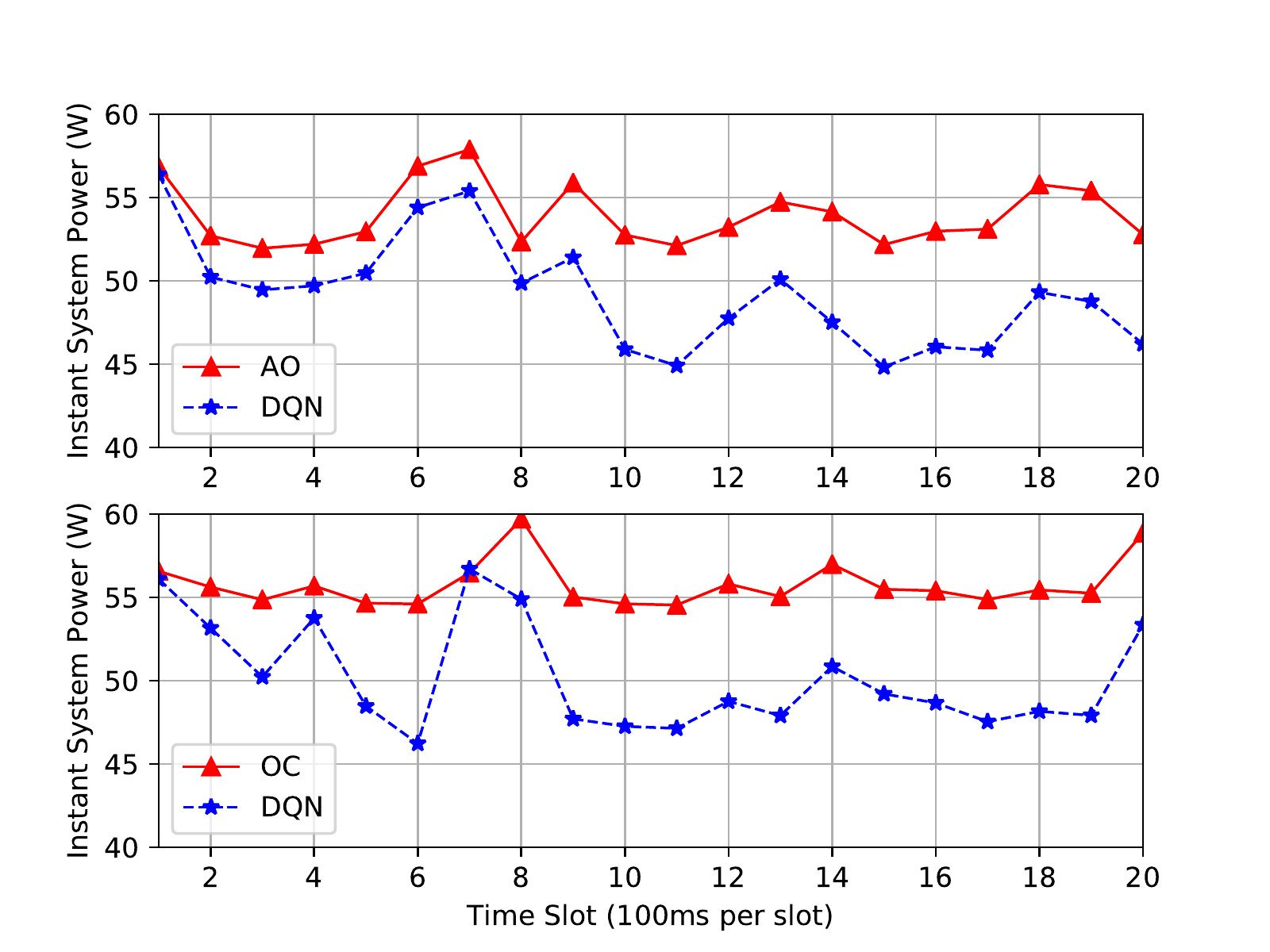}}
    \label{2b}
	  \caption{Total instant power consumption with different user demands and allocation schemes.}
	  \label{fig4} 
\end{figure}

We examine the instant system power consumption in this subsection. In the top figures of Fig.~\ref{fig4}(a) and Fig.~\ref{fig4}(b), we compare the strategies of AO and DQN, where we set all the RRHs open initially and then all RRHs stay active in AO schemes. In the bottom figures of Fig.~\ref{fig4}(a) and Fig.~\ref{fig4}(b), we turn off one RRH randomly at the beginning for both OC and DQN and then one RRH stay switched off in OC scheme. Moreover, we set user demands are selected randomly from the set of 20\textit{Mbps} to 40\textit{Mbps} in Fig.~\ref{fig4}(a), whereas we randomly select user demands from the set of 20\textit{Mbps} to 60\textit{Mbps} in Fig.~\ref{fig4}(b). One can see from all the figures in Fig.~\ref{fig4} that our proposed DQN always outperforms AO and OC. This is because DQN controls RRHs to turn on and off depending on the current states of the systems, whereas AO always turns on all the RRHs and OC randomly turns off one RRH, which may not be the optimal strategy and contribute to larger power consumption than DQN.

One can also see that when we increase the upper limit of user demands from
40\textit{Mbps} in Fig.~\ref{fig4}(a) to 60\textit{Mbps} in Fig.~\ref{fig4}(b), the performance of all DQN, OC and AO become more unstable. However, our proposed DQN still has the best performance when compared with AO and OC.

\begin{figure} 
    \centering
	  \subfloat[AO VS. DQN]{
       \includegraphics[width=3.5 in]{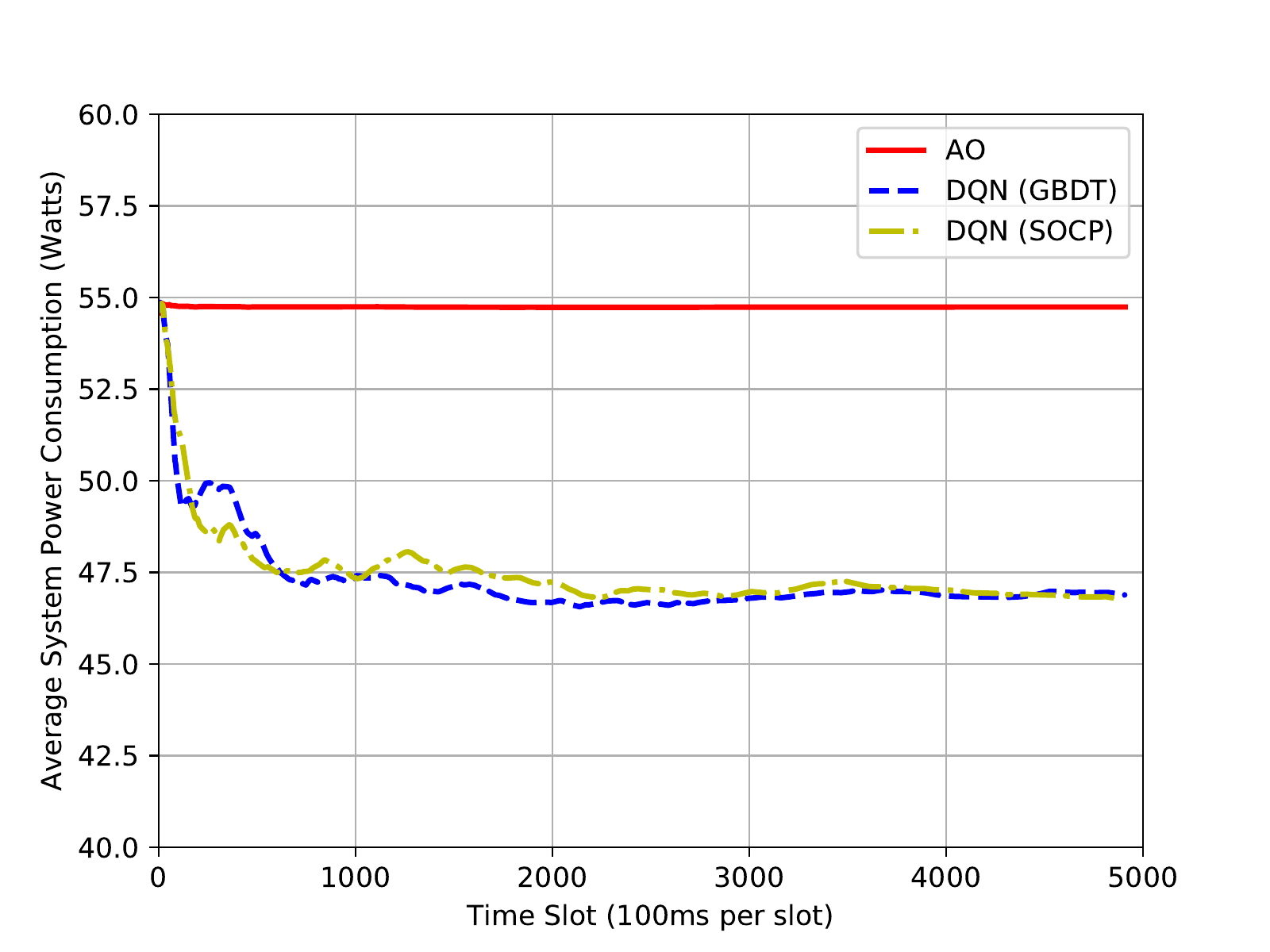}}
    \label{3a}\\
	  \subfloat[OC VS. DQN]{
        \includegraphics[width=3.5 in]{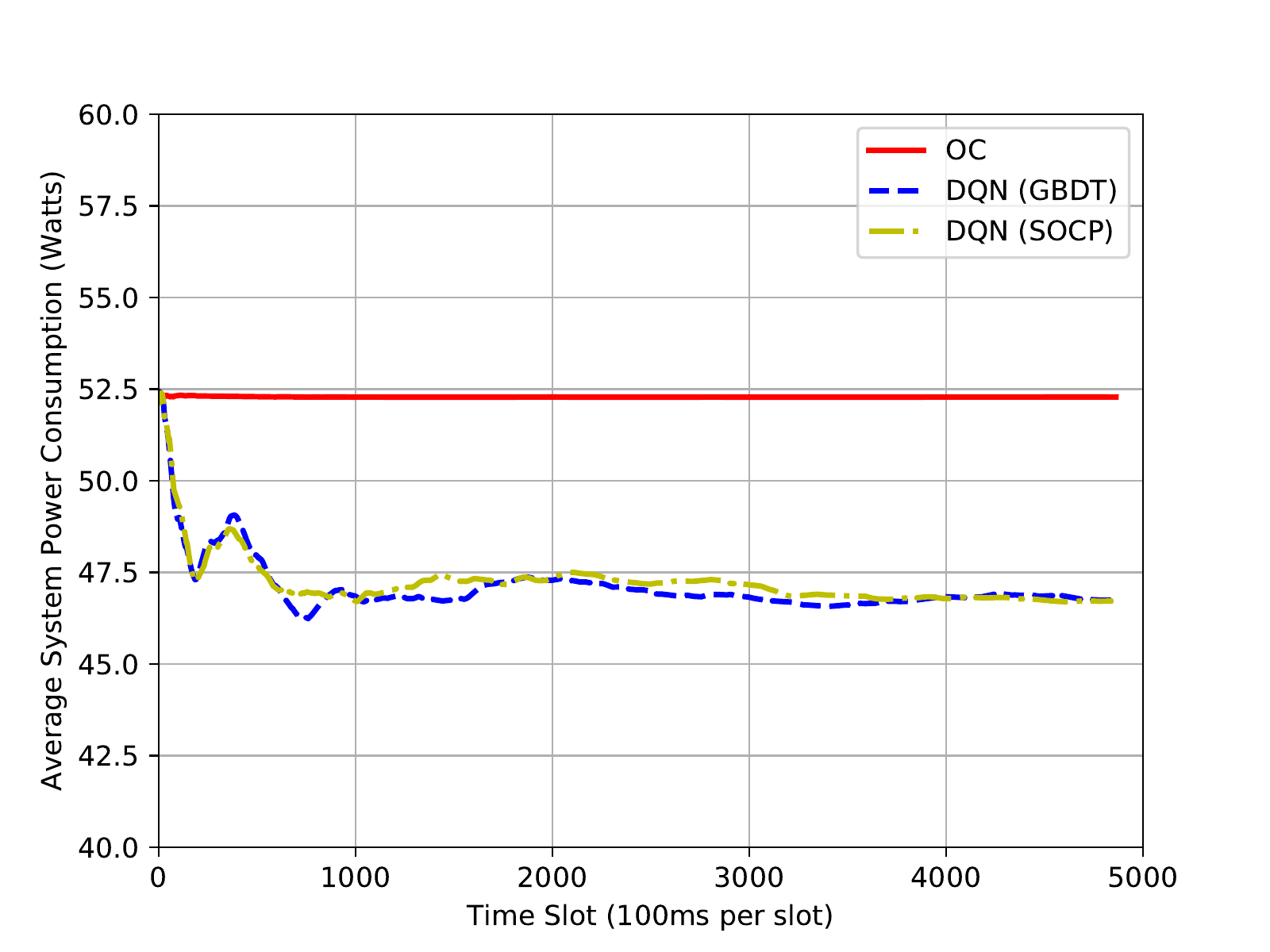}}
    \label{3b}
	  \caption{Total average power consumption with different allocation schemes.}
	  \label{fig5} 
\end{figure}

Moreover, one can see that although there may be some errors caused by GBDT approximator, our proposed DQN framework has considerable performance, which shows the good ability of error tolerance in our proposed solution.

\subsubsection{Average Power} 
In Fig.~\ref{fig5}, we show the performance comparison between GBDT-based DQN, AO and OC in the long term. The DQN with reward obtained from SOCP solver is also depicted. We compare the average system power consumption by averaging all instant system power in the past time slots. 

We first analyse the performance under the condition of user demands below 40\textit{Mbps} between both DQN schemes (including GBDT and SOCP) and AO scheme. 
We set all the RRH switched on and set user demands changed every 100 ms per slot and lasting for 500\textit{s}. One can see from Fig.~\ref{fig5}(a) that both DQN schemes outperform AO and can save power around 8 Watts per time slot. The slight fluctuation comes from the randomness of the requirement. Moreover, one can see from Fig.~\ref{fig5}(a) that DQN with GBDT have the similar performance as the DQN scheme with SOCP solver, which shows the error tolerance feature of our proposed solutions. 


Then we turn one RRH off and continue to analyse the average system power consumption under DQN and OC scheme. One can see from Fig.~\ref{fig5}(b) that both DQN schemes still outperform OC scheme, as expected. Also, one can see that DQN scheme with GBDT has the similar performance as SOCP solver, similarly with above. 

\begin{figure} 
    \centering
	  \subfloat[AO VS. DQN]{
       \includegraphics[width=3.5 in]{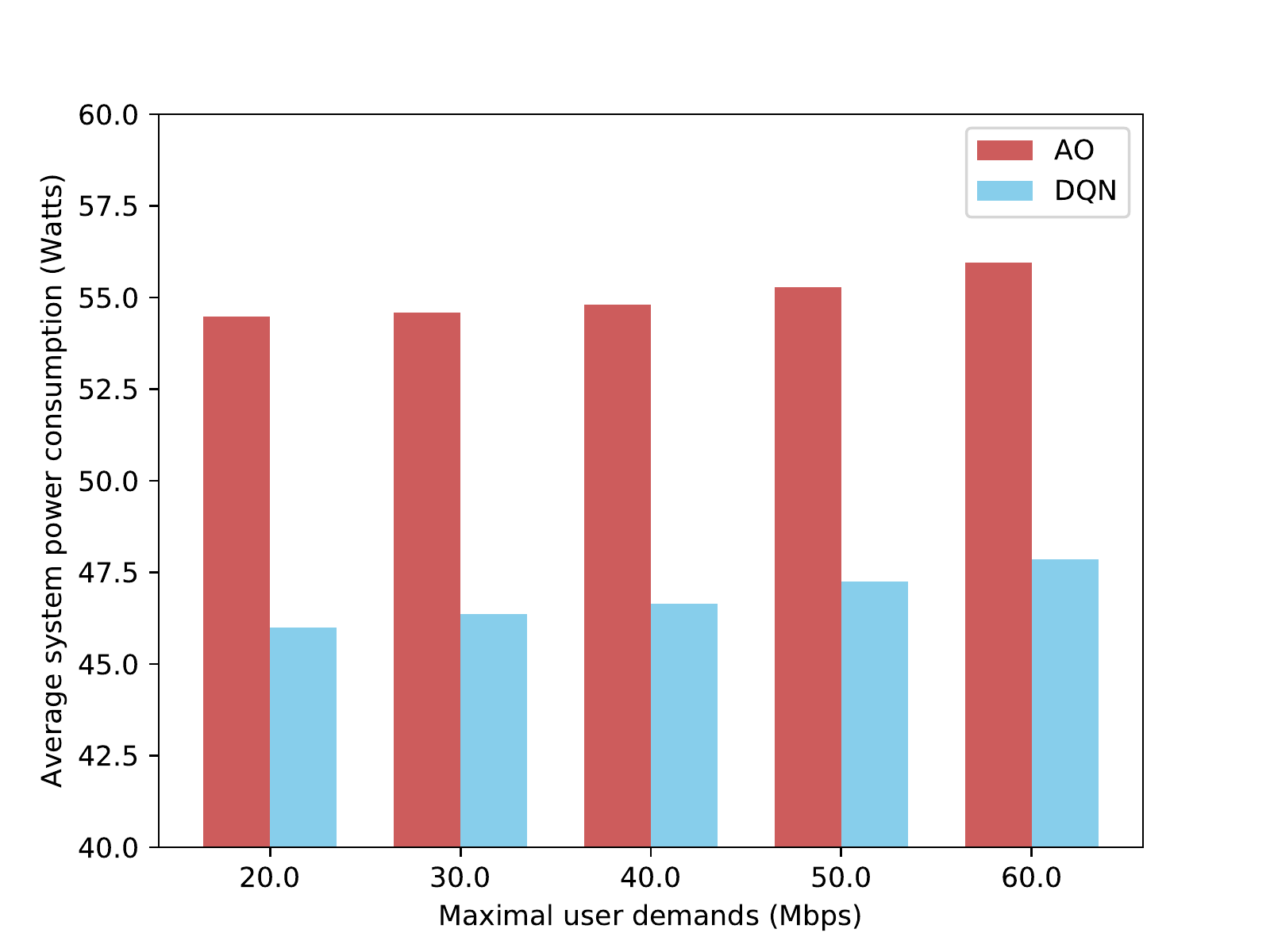}}
    \label{6a}\\
	  \subfloat[OC VS. DQN]{
        \includegraphics[width=3.5 in]{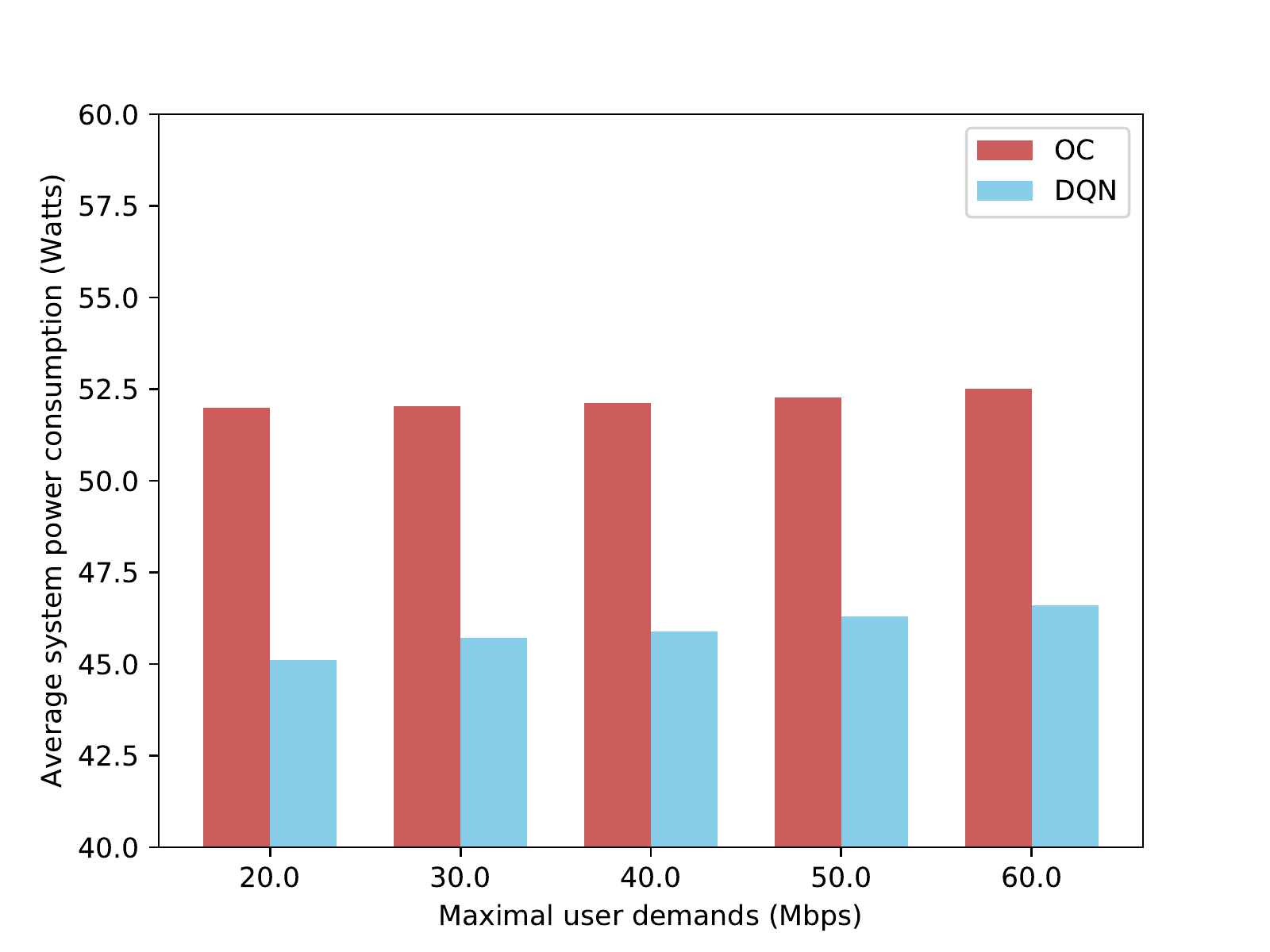}}
    \label{6b}
	  \caption{Average power consumption versus different user demands.}
	  \label{fig6} 
\end{figure}


\subsubsection{Overall Performance of GBDT-based DQN} 
To evaluate the overall performance of GBDT-based DQN in different situations, we set user demands from 20\textit{Mbps} to 60\textit{Mbps} with 10\textit{Mbps} interval, and keep other factors unchanged. One can see from Fig.~\ref{fig6}(a) and Fig.~\ref{fig6}(b) that with the increase of user demands, the power consumption of AO, OC and DQN increase as well. One also sees that our proposed GBDT-based DQN have much better performance than AO and OC, as expected, which prove the effectiveness of our scheme.

 


\section{Conclusion}
In this paper, we presented a GBDT-based DQN framework to tackle the dynamic resource allocation problem for IoT in the real-time C-RANs. 
We first employed the GBDT to approximate the solutions of the SOCP problem. Then, we built the DQN framework to generate a efficient resource allocation policy regarding to the status of RRHs in C-RANs. Furthermore, we demonstrated the offline training, online decision making as well as regular tuning processes.
Lastly, we evaluated the proposed framework with the comparison to two other methods, AO and OC, and examined its accuracy and the ability of error tolerance compared with SOCP-based DQN scheme.
Simulation results showed that the proposed GBDT-based DQN can achieve a much better performance in terms of power saving than other baseline solutions under the real-time setting. Future work is in progress to let GBDT approximator meet the strict constraints of practical problems, which is expected to be employed in a wide range of scenarios.

\appendix[Training and Predicting Process of GBDT]

The training process of GBDT is shown in Algorithm 2.

\begin{algorithm}
\caption{Training process of GBDT}
\begin{algorithmic}
\State{\textbf{Initialization}}
\State{(1) Set the iteration counter $m=0$. Initialize the additive predictor $\widehat{f}^{[0]}$ with a starting value, e.g. $\widehat{f}^{[0]}:=(0)_{i=1,\cdots,n}.$ Specify a set of base-learners $h_1 (x_1),\cdots,h_p (x_p)$}
\State{\textbf{Fit the negative gradient}}
\State{(2) Set $m:=m+1$}
\State{(3) Compute the negative gradient vector $\textbf{u}$ of the loss} 
\State{\qquad function at the previous iteration:}
\begin{equation*}
\textbf{u}^{|m|}=(u_i^{|m|})_{i=1,\cdots,n}=({-\frac{\partial}{\partial f}\rho(y_i,f)}|_{f=\widehat{f}^{|m-1|}(\cdot)})_{i=1,\cdots,n}
\end{equation*}

\State{(4) Fit the negative gradient vector $\textbf{u}^{|m|}$ separately to every} 
\State{\qquad base-learner:}
\begin{equation*}
\textbf{u}^{|m|} \xrightarrow{base-learner} \widehat{h}_j^{[m]}(x_j) \ \text{for} \ j=1,\cdots,p    
\end{equation*}
\State{\textbf{Update one component}}
\State{(5) Select the component $j^{*}$ that best fits the negative} 
\State{\qquad gradient vector:}
\begin{equation*}
j^{*}= \arg\min \limits_{1\leq j \leq p} \sum\limits_{i=1}^n (u_i^{|m|}-\widehat{h}_j^{|m|}(x_j))^2  
\end{equation*}
\State{(6) Update the additive predictor $\widehat{f}$ with the component}
\begin{equation*}
\widehat{f}^{[m]}(\cdot) = \widehat{f}^{[m-1]}(\cdot) + sl \cdot \widehat{h}_{j^*}^{|m|}(x_{j^*})
\end{equation*}
\State{\qquad where $sl$ is a small step length $(0< sl \ll 1$ and a }
\State{\qquad typical value in practice is 0.1.}
\State{\textbf{Iteration}}
\State{Iterate steps (2) to (6) until $m=m_{stop}$}
\label{GBDT}
\end{algorithmic}
\end{algorithm}

The GBDT is consisted of two concepts, where one is called the gradient and the other is boosting. 
In training process, the 0-th tree is fitted to the given training dataset, and it predicts the mean value of $y_{\text{true}}$ in the training set regardless of what the input is; the predicted values of 0-th tree are denoted as $y_{\text{predicted}_0}$. However, the predictions $y_{\text{predicted}_0}$ from the 0-th tree still have residuals between true values $y_{\text{true}}$. Then, another additive tree is applied to fit to the new dataset that the inputs are same as the 0-th tree, but the fitting target $y$'s are the residuals $(y_{\text{true}}-y_{\text{predicted}_0})$. Then, the predictions of the GBDT are the linear combination of the predictions from the 0-th tree and the new additive tree, namely $y_{\text{predicted}}=y_{\text{predicted}_0}+\gamma_1*y_{\text{predicted}_1}$, where $\gamma_1$ is the weight attributed to this tree. Next, another tree is fitted to the new residuals $y_{\text{true}}-(y_{\text{predicted}_0}+\gamma_1*y_{\text{predicted}_1})$ and follow the same process as before.

From above process, one can see that the boosting concept is to utilize the residuals between the previous ensembled results and true values. By learning from the residual, the model can make progress when new trees are added. The gradient part of concept can be explained as that the whole training process is supervised and guided by the gradient of objective function, where it is typically expressed as $0.5*(y_{\text{true}}-y_{\text{predicted}})^2$, whose derivative is the pseudo-residual between $y_{\text{true}}$ and $y_{\text{predicted}}$. 

\ifCLASSOPTIONcaptionsoff
  \newpage
\fi

\end{document}